\documentclass[journal,9pt]{IEEEtran}

\usepackage{amsthm, amssymb}
\usepackage{amsmath}
\usepackage{comment}
\usepackage{graphicx}
\usepackage{cleveref}
\usepackage{amssymb}
\usepackage{booktabs}
\usepackage{subcaption}

\usepackage[style=ieee,backend=biber, url=false, doi=false, isbn=false]{biblatex}
\Crefname{figure}{Fig.}{Fig.}

\ifCLASSINFOpdf
\else
\fi
\begin{document}
%
\title{Homotopy-Based Re-Initialization for Switched DAEs in Power System Transient Simulation}

\author{Ahmad~Ali,~\IEEEmembership{Graduate Student Member,~IEEE,}
        Hantao~Cui,~\IEEEmembership{Senior Member,~IEEE}
}

\markboth{Journal of \LaTeX\ Class Files,~Vol.~14, No.~8, August~2015}%
{Shell \MakeLowercase{\textit{et al.}}: Bare Demo of IEEEtran.cls for IEEE Journals}

\maketitle

\begin{abstract}
The simultaneous solution of switched differential-algebraic equations (DAEs) in power system transient simulation may suffer convergence loss following discontinuous events.
This difficulty is typically interpreted as a poor post-event initialization problem.
This letter presents a geometric framework that explains the underlying convergence mechanism and clarifies why standard convergence-restoration methods may fail at discontinuities.
Based on this interpretation, a homotopy-continuation based globalized re-initialization scheme is developed to restore convergence.
The proposed method is validated through numerical simulations of representative discontinuities in power system transient simulation.
Results show that in the cases where direct post-event solution fails, the proposed scheme can reliably recover convergence.
\end{abstract}

\IEEEpeerreviewmaketitle

\section{Introduction}
Power system transient simulation is commonly formulated as the numerical solution of switched differential-algebraic equations (DAEs) \cite{milanoPowerSystemModelling2010a}.
Under discontinuous events such as breaker operations, faults, and controller limiting, the simultaneous solution 
of these DAEs may suffer convergence loss in the nonlinear solve \cite{laraRevisitingPowerSystems2023}.
In particular, Newton-Raphson (NR) iterations may stagnate or diverge immediately following a discontinuous event even when a valid post-event solution exists.
 
Mathematically, power system switched DAEs are expressed as: 
\begin{equation} 
\label{eq:index1_dae}
    \dot{\mathbf{x}} = \mathbf{f}_{\sigma}(\mathbf{x}, \mathbf{y})\, ; \quad   0 = \mathbf{g}_{\sigma}(\mathbf{x}, \mathbf{y})
\end{equation}
where $\mathbf{x}$ and $\mathbf{y}$ are the differential and algebraic variables, respectively, $\mathbf{f}$ and $\mathbf{g}$ represent the differential and algebraic equations, respectively, and $\sigma \in S$ denotes the discreet system mode.
A change in $\sigma$ replaces one set of continuous equations with another, thereby introducing a switched DAE.

For continuous systems, integration step size control is a standard approach to improve nonlinear solver convergence \cite{soderlindAdaptiveTimesteppingComputational2006}, since the solution from the previous time step provides an increasingly accurate initial guess as the step size decreases.
However, this approach may fail for switched DAEs because the algebraic variables can undergo discontinuous changes at mode transitions.
Consequently, the pre-event state may remain numerically distant from the post-event
solution even under aggressive step size reduction.

The other existing solution is algebraic re-initialization \cite{brownConsistentInitialCondition1998}, in which the algebraic equations are re-solved at the mode transition with the differential variables held fixed.
Although globally convergent Newton methods 
can 
improve robustness, such methods exhibit fundamental limitations in root-finding problems.
In particular, these methods typically minimize a residual-based merit function
and may converge to a stationary point that is not a root of the original system.
A more robust re-initialization strategy is therefore needed.

This letter addresses the question: why does convergence fail after a mode transition even when a valid post-event solution exists?
The main premise is that the difficulty is not merely numerical, but geometric.
By interpreting the algebraic equations as defining constraint manifolds in the state space, this letter shows that a mode transition changes the solution manifold, rendering the pre-event state inconsistent with the post-event system. Based on this interpretation, a homotopy-continuation based re-initialization scheme is proposed to compute a consistent post-event point and restore convergence.

The main contribution of this letter is a homotopy-continuation-based re-initialization scheme for switched power system DAEs that restores nonlinear solver convergence at discontinuous events where direct post-event solution and step-size reduction fail. A geometric interpretation of the underlying mechanism is provided to motivate the approach and clarify why standard remedies are insufficient.

This letter is organized as follows: 
Section II presents the geometric analysis of switched DAEs.
Section III introduces the HC-based re-initialization scheme.
Section IV presents case studies and simulation results, followed by
conclusion in section V.

\section{Geometric Interpretation of Switched DAEs}
This section analyzes the geometric structure of the
switched DAE in \eqref{eq:index1_dae}.
For a system mode $\sigma$, the governing algebraic equations define a manifold $\mathcal{M_{\sigma}}$ in the state space:
\begin{align}
    \mathcal{M_{\sigma}} := \left \{(x,y) \in \mathbb{R}^{n_x + n_y} \; \middle| \; g_{\sigma}(x,y) = 0\ \right \}
\end{align}
DAE velocity vector $v$, which describes the evolution of system trajectories in the state space, is defined as:
\begin{align}
    v = (\dot{x}, \dot{y})
\end{align}
where $\dot{x}$ and $\dot{y}$ are the time derivatives of the state and algebraic variables, respectively.

To characterize the relation between the manifold and trajectory evolution, differentiate the algebraic equation with respect to time:
\begin{align}
    g_{\sigma, x} \dot{x} + g_{\sigma, y} \dot{y} &= 0
    \label{eq:tangency_condition_initial}
\end{align}
where $g_{\sigma, x} = \partial g_{\sigma}/\partial x$ and $g_{\sigma, y} = \partial g_{\sigma}/\partial y$
are the Jacobian matrices of $g_{\sigma}$ with respect to $x$ and $y$, respectively.
Equivalently,
\begin{align}
    [\,g_{\sigma,x} \;\; g_{\sigma,y}\,]\, v = 0
    \label{eq:tangency_condition_final}
\end{align}
Since the row space of the constraint Jacobian $[\,g_{\sigma, x} \;\; g_{\sigma, y}\,]$ defines the normal space to the manifold at $(x,y)$, the orthogonality condition in \eqref{eq:tangency_condition_final}
implies that $v$ lies in the tangent space of $\mathcal{M}_{\sigma}$.
Therefore, the DAE vector field is tangent to the solution manifold.

This implies that the algebraic equations constrain the state space 
such that solution trajectories flow on the manifold defined by these equations. 
Provided that $\partial g_{\sigma}/\partial y$ is non-singular,
\eqref{eq:tangency_condition_initial} yields
\begin{align}
    \dot{y} &= -g_{\sigma, y}^{-1} \; g_{\sigma, x}\; \dot{x}
\end{align}
Then, by the implicit function theorem (IFT), there exists a local map $y=\phi_{\sigma}(x)$ such that the DAE can be reduced locally to an ODE $\dot{x} = f_{\sigma}(x, \phi_{\sigma}(x))$.
With an appropriate choice of state variables, power system transient models are typically formulated as index-1 DAEs 
\cite{laraRevisitingPowerSystems2023},
for which $g_{y}$ is non-singular.
Hence, a switched power system DAE can be interpreted as an ODE evolving on a manifold.

This interpretation also clarifies the meaning of a consistent initial condition.
Let $z:=(x, y)$ denote a point in the state space.
This point is \emph{consistent} if it satisfies the algebraic constraints of the active mode, i.e., $g_{\sigma}(z) = 0$,
so that $z \in \mathcal{M_{\sigma}}$.
Such a point defines a locally admissible initial condition.
Under the regularity conditions of the IFT, a unique local solution to \eqref{eq:index1_dae} exists, and the corresponding trajectory evolves on the manifold $\mathcal{M}_{\sigma}$ \cite{hairerSolvingDifferentialEquations}.

Now consider a mode transition at $t=t_s$ that changes the system mode from $\sigma^{-}$ to $\sigma^{+}$.
In the simultaneous solution approach, an implicit integration method advances the solution by solving a nonlinear system of the form $R_{\sigma}(z_{k+1}; z_k, h) = 0$ at a time step $k$, 
where $h$ is the integration step size, 
and $R_{\sigma}$ is the residual associated with the active mode.
For example, using the implicit Euler method,
\begin{align}
\label{eq:be_residual}
R_{\sigma}(z_{k+1}; z_k, h) =
\begin{pmatrix}
    x_{k+1} - x_{k} - hf_{\sigma}(x_{k+1}, y_{k+1}) \\
    g_{\sigma}(x_{k+1}, y_{k+1})
\end{pmatrix}
\end{align}

A mode transition changes the admissible constraint manifold, so that the pre-event solution generally becomes inconsistent with the new post-event manifold.
Immediately before the event, 
$z_{s}^{-} := (x(t_{s}^{-}), y(t_{s}^{-})) \in \mathcal{M_{\sigma^{-}}}$.
However, after the transition, the algebraic equations are replaced by those of the new mode, and in general, $g_{\sigma^{+}}(z_{s}^{-}) \neq 0$.
Hence, $z_{s}^{-} \notin \mathcal{M_{\sigma^{+}}}$.

The post-event nonlinear system is then solved iteratively using the NR method:
\begin{align}
    z_{k+1}^{(m+1)} = z_{k+1}^{(m)} - \left[J_{\sigma^+}\bigl(z_{k+1}^{(m)}\bigr)\right]^{-1} R_{\sigma^+}\bigl(z_{k+1}^{(m)}; z_k, h\bigr)
    \label{eq:nr_iteration}
\end{align}
where $m$ denotes the NR iteration number, and $J_{\sigma^{+}}$ is the Jacobian of $R_{\sigma^{+}}$ with respect to $z_{k+1}$.
At the first post-event step, the initial NR iterate is typically chosen from the previous solution i.e. $z_{k+1}^{(0)} = z_{s}^{-}$, which is inconsistent with the post-event manifold.

Since the DAE dynamics are admissible only on the constraint manifold, no valid post-event DAE trajectory exists that continuously connects an inconsistent point to $\mathcal{M_{\sigma^{+}}}$.
Therefore, at the first post-event step, the solver must recover a consistent point on $\mathcal{M}_{\sigma^{+}}$ before integration can proceed.
This recovery cannot be achieved through finite flow along the post-event vector field, because the vector field is tangent to $\mathcal{M_{\sigma^{+}}}$ and is not defined away from it.
As a result, direct post-event integration may fail to converge.

Numerically, consistency restoration requires an instantaneous correction of the algebraic variables i.e. $\dot{y}\rightarrow\infty$.
The NR method is therefore forced to resolve this impulsive transient which can severely degrade the conditioning of the NR step and cause the iterations
to stagnate or diverge.
Moreover, the NR method is only locally convergent.
Under strong nonlinearities, the inconsistent point may
lie outside the basin of convergence of the post-event solution.
In that case, the NR iterations may again stagnate or diverge.

This geometric mismatch  
also explains why reducing the step size alone cannot resolve the difficulty.
From \eqref{eq:be_residual}, decreasing $h$ scales only the differential part of the residual, while the algebraic inconsistency with respect to $g_{\sigma^{+}}$ remains unchanged.
Hence, step size reduction may reduce the discretization error associated with the continuous time dynamics, it does not restore consistency with the post-event manifold.
A variable step solver may therefore continue to shrink the step size without recovering convergence at the discontinuity.

\begin{figure}\label{fig: main_idea}
    \centering
    \includegraphics[width=0.5\linewidth]{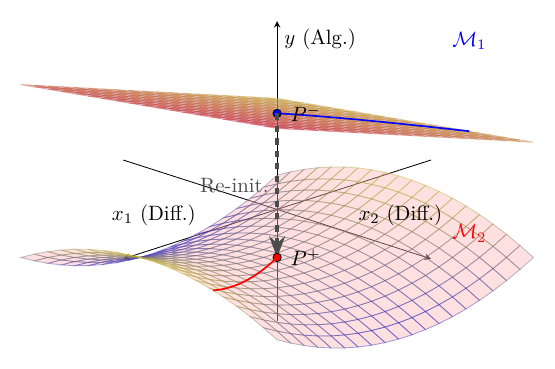}
    \caption{Geometric illustration of mode transition in switched DAEs.}
    \label{fig:main_idea_pic}
\end{figure}

\Cref{fig:main_idea_pic} depicts DAE trajectory evolution on a manifold, and the inconsistency of a solution following mode transition.

\section{Homotopy-Based Re-initialization}
The re-initialization problem consists of finding $z_s^{+} \in M_{\sigma^{+}}$ given $z_s^{-} \in M_{\sigma^{-}}$.
The proposed approach builds on Brown's re-initialization scheme \cite{brownConsistentInitialCondition1998}.
Specifically, with the pre-event differential variables $x_s^-$ held fixed, one seeks $y_s^+$ such that
\begin{align}
    g_{\sigma^{+}}(x_s^-, y_s^+) = 0
\end{align}
This yields the post-event point $(x_s^-, y_s^+)$ which is consistent with the new mode.
However, this re-initialization step is itself a nonlinear solve, typically carried out using NR-like iterations.
Consequently, its success still depends on whether the inherited point lies within the region of convergence of the post-event solution.

To improve the robustness of this step,
this work proposes a homotopy-continuation (HC)-based globalized re-initialization strategy.
Let $\mathcal{X}$ denote the state space, with $M_{\sigma^{-}}, M_{\sigma^{+}} \subset \mathcal{X}$ 
The proposed method constructs a continuous path 
$\gamma:[0, 1] \rightarrow \mathcal{X}$, such that $\gamma(0) = z_s^{-}$ and $\gamma(1) = z_s^{+}$.
This is achieved by introducing a family of intermediate manifolds that gradually connects the pre-event solution to the post-event manifold.
This replaces the original nonlinear solve by a sequence of nearby sub-problems.
Since consecutive continuation
problems remain close, the NR method is more likely to
remain convergent than in a direct one-shot re-initialization.

Mathematically, the HC-based re-initialization is posed as:
\begin{align}
    H(z ; \lambda)=0, \qquad \lambda\in[0,1]
    \label{eq:homotopy_general}
\end{align}
where $z$ denotes the variables to be initialized and $\lambda$ is the continuation parameter. The homotopy is constructed so that $H(z ; 1)=g_{\sigma^+}(z)$, 
while the initial system, $H(z ; 0) = H_0(z)$, 
is an auxiliary problem with a known solution.
In the simplest case, $H_0$ may be chosen as the pre-event algebraic system.
More generally, $H_0$ may be defined as a modified version of the post-event system such that the inherited pre-event point is an exact solution at $\lambda=0$. 
Continuation in $\lambda$ then traces a sequence of nearby nonlinear problems from the known start solution to a consistent post-event solution. 

After continuation step $j$, $\lambda$ is updated as 
$\lambda_{j+1} = \lambda_j + \Delta \lambda_j$, 
and the nonlinear system, $H(z_{j+1}; \lambda_{j+1}) = 0$,
is solved for $z_{j+1}$ using the converged solution from the previous continuation step as the initial guess.
To improve efficiency, the continuation increment, $\Delta \lambda$, can be selected adaptively.
This allows larger steps when the intermediate nonlinear solves are well-behaved and smaller steps when convergence becomes more difficult.

The proposed scheme differs from prior homotopy-based work.
Homotopy continuation has been applied to DAE cold-start initialization and to ill-conditioned power flow, but in both cases, the homotopy connects an auxiliary problem to a single target system solved once.
In contrast, the proposed scheme targets \emph{in-simulation} re-initialization at mode transitions: it connects the pre-event manifold $\mathcal{M}_{\sigma^-}$ to the post-event manifold $\mathcal{M}_{\sigma^+}$ through a family of intermediate manifolds, with $H_0$ inheriting the pre-event point $z_s^-$ as an exact solution at $\lambda=0$.
This extends Brown's single-step re-initialization \cite{brownConsistentInitialCondition1998} with a globalization mechanism to guarantee a valid starting iterate, regardless of the distance between $\mathcal{M}_{\sigma^-}$ and $\mathcal{M}_{\sigma^+}$.

\section{Case Studies}
Two case studies are presented.
The first considers a circuit where a mode transition places the inherited point outside the basin of attraction of the post-event solution.
The second examines a network topology change that causes divergence of the nonlinear solver.

The DAE system is integrated using the implicit trapezoidal method.
In the nonlinear solve \eqref{eq:nr_iteration}, the Jacobian is updated at every iteration to avoid potential convergence degradation.
In all cases reported below, a valid post-event solution exists.
\subsection{Case Study 1: Control Mode Switching}
This case study illustrates loss of NR convergence following a manifold jump induced by control-mode switching in a grid-forming (GFM) converter.
For simplicity, a reduced steady-state converter model is used.
Under normal operation, the GFM converter acts as a voltage source;
however, when the converter output current reaches its maximum limit, the control mode transitions to current limiting operation.
A constant-power load of magnitude $P$ is connected at the converter terminals.

In the GFM mode:
\begin{align}
    g_{\sigma^{-}}(V) = \frac{V_{s} - V}{R} - \frac{P}{V} = 0 
    \label{eq:gfm}
\end{align}
where $V_s$ is the converter voltage, $V$ is the terminal voltage, and $R$ is the line resistance.
In current limiting mode,
\begin{align}
    g_{\sigma^{+}}(V) = I_{\max} - \frac{P}{V} = 0 
    \label{eq:gfl}
\end{align}
where $I_{\max}$ denotes the converter current limit. 
%
Since the constraint manifold is determined by the algebraic equations of the active mode, the mode transition induces manifold switch.
 
\begin{figure}
    \centering
    \includegraphics[width=1.0\linewidth]{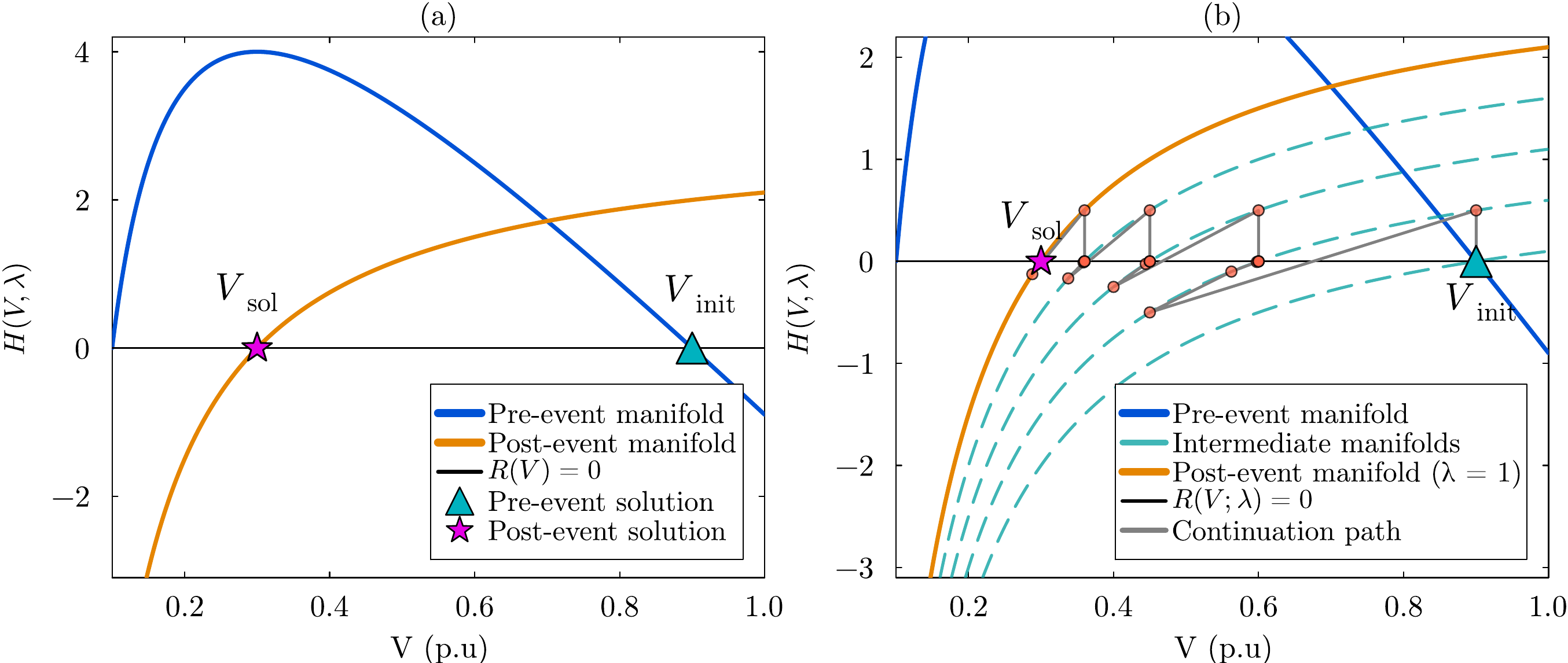}
    \caption{(a) Solution manifolds for GFM and current-limiting modes. (b) HC-based re-initialization process.}
    \label{fig:converter_mode_switch}
\end{figure}

\begin{figure}
    \centering
    \includegraphics[width=1.0\linewidth]{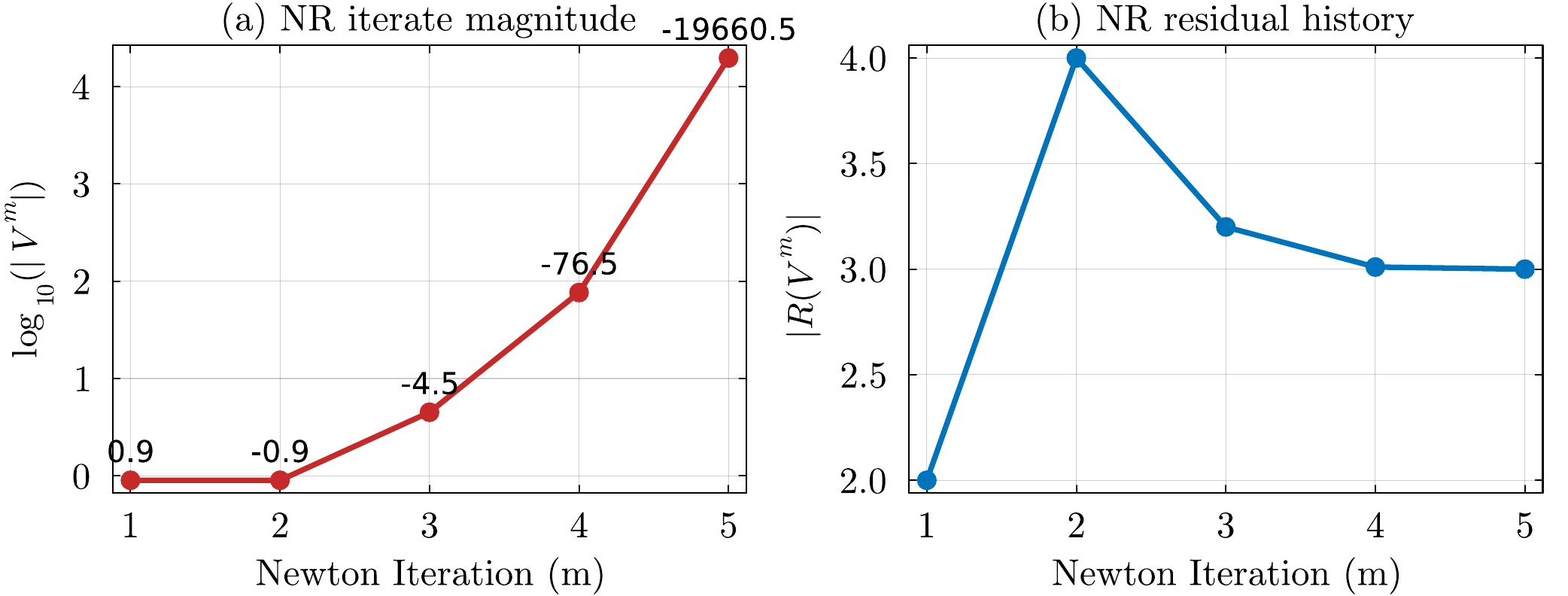}
    \caption{NR convergence behavior without HC.}
    \label{fig:converter_divergence}
\end{figure}

At mode transition, direct NR iterations fail to converge.
\Cref{fig:converter_mode_switch}(a) shows the manifolds defined by \eqref{eq:gfm} and \eqref{eq:gfl}, together with the solution points before and after the mode transition.
\Cref{fig:converter_divergence} shows the behavior of the NR iterations when the pre-event point $V_{init}$ is used to solve \eqref{eq:gfl}.
\Cref{fig:converter_divergence}(a) and (b)
show the logarithmic magnitude of the NR iterates and the residual norm, respectively, as the iterations proceed.
The residual stagnates and fails to convergence. 

Using the proposed scheme, the solver converges to the desired solution.
\Cref{fig:converter_mode_switch}(b) illustrates the continuation process.
The pre-event and post-event manifolds are connected through a sequence of intermediate manifolds,
such that the solution from each continuation step remains within the basin of attraction of the next subproblem.

\subsection{Case Study 2: Topology Change}
This case study examines convergence loss induced by a topology change that alters the post-event constraint manifold.
A three-phase line-to-ground fault is simulated on the modified IEEE 39-bus system.
At $t=1.0 s$, a fault impedance of $Z_f=j0.015$ is connected at bus 20.
Solver behavior is examined at the instant of fault application.
The system is modeled in electromagnetic transient detail using the state-space formulation and solved with the simultaneous approach.

\begin{figure}
    \centering
    \includegraphics[width=1.0\linewidth]{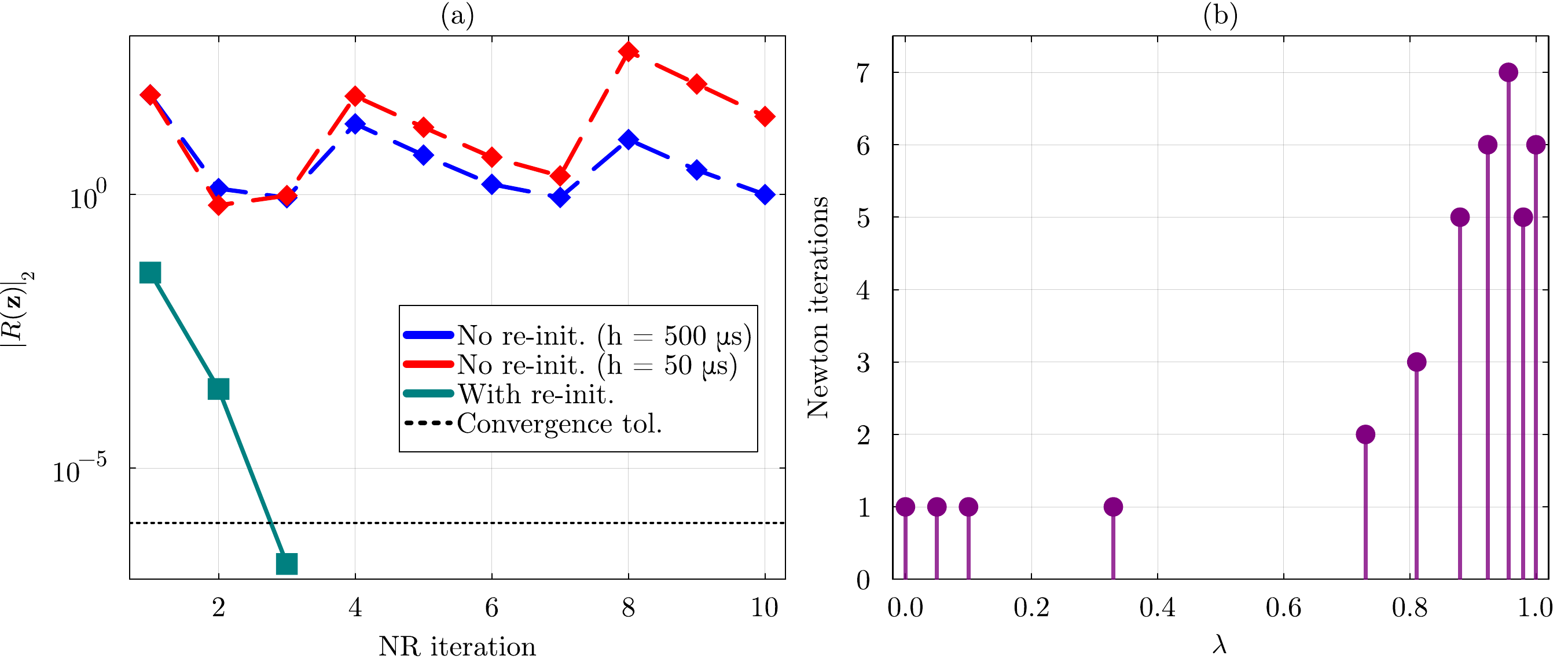}
    \caption{(a) NR convergence behavior with and without HC-based re-init. (b) NR iteration count for intermediate continuation problems.}
    \label{fig:39bus16_nr}
\end{figure}

Without re-initialization, the NR iterations fail to converge even when the integration step size is reduced, as shown in \cref{fig:39bus16_nr}(a).
The NR iterations stagnate and the residual norm remains well above the convergence tolerance.
This behavior is observed for both tested integration step sizes, $500~\mu s$ and $50~\mu s$.
Reducing the step size by one order of magnitude therefore does not resolve the post-event convergence difficulty. 
This supports the geometric interpretation developed earlier: the dominant issue is loss of consistency with the post-event algebraic manifold, rather than discretization error.

For the same system, however, the proposed HC-based re-initialization scheme successfully computes a consistent post-event point.
When this point is used to initialize the first post-event nonlinear solve, the solver converges in three iterations, as shown in \cref{fig:39bus16_nr}(a).
The much smaller residual at the first NR iteration in the re-initialized case indicates that the nonlinear solve begins substantially closer to the desired solution.
\Cref{fig:39bus16_nr}(b) 
shows the NR iterations for the intermediate continuation problems during the re-initialization stage.

\section{Conclusion}
This letter presents a geometric 
interpretation of
convergence issues in the simultaneous solution of switched DAEs, and 
develops a robust homotopy-continuation based re-initialization scheme for discontinuous events.
The key observation is that a mode transition changes the solution manifold, rendering the pre-event solution inconsistent with the post-event manifold.
Convergence difficulty therefore stems from the nonexistence of a valid continuous trajectory from the inherited point to the post-event manifold.
By introducing a sequence of intermediate manifolds, the proposed scheme computes a
consistent post-event initialization robustly.
Its effectiveness is demonstrated in two different manifold-switching scenarios where direct post-event solution fails but the proposed scheme restores convergence.

\ifCLASSOPTIONcaptionsoff
  \newpage
\fi

\printbibliography

@article{brownConsistentInitialCondition1998,
  title = {Consistent Initial Condition Calculation for Differential-Algebraic Systems},
  author = {Brown, Peter N and Hindmarsh, Alan C and Petzold, Linda R},
  date = {1998},
  journaltitle = {SIAM Journal on Sci. Comp.},
  volume = {19},
  number = {5},
  pages = {1495--1512},
  publisher = {SIAM}
}

@article{hairerSolvingDifferentialEquations,
  title = {Solving {{Differential Equations}} on {{Manifolds}}},
  author = {Hairer, Ernst},
  langid = {english}
}

@article{laraRevisitingPowerSystems2023,
  title = {Revisiting Power Systems Time-Domain Simulation Methods and Models},
  author = {Lara, Jose Daniel and Henriquez-Auba, Rodrigo and Ramasubramanian, Deepak and Dhople, Sairaj and Callaway, Duncan S and Sanders, Seth},
  date = {2023},
  journaltitle = {IEEE Transactions on Power systems},
  volume = {39},
  number = {2},
  pages = {2421--2437},
  publisher = {IEEE}
}

@book{milanoPowerSystemModelling2010a,
  title = {Power {{System Modelling}} and {{Scripting}}},
  author = {Milano, Federico},
  date = {2010},
  series = {Power {{Systems}}},
  volume = {0},
  publisher = {Springer Berlin Heidelberg},
  location = {Berlin, Heidelberg},
  doi = {10.1007/978-3-642-13669-6},
  url = {http://link.springer.com/10.1007/978-3-642-13669-6},
  urldate = {2026-04-14},
  isbn = {978-3-642-13668-9 978-3-642-13669-6},
  langid = {english}
}

@article{soderlindAdaptiveTimesteppingComputational2006,
  title = {Adaptive Time-Stepping and Computational Stability},
  author = {Söderlind, Gustaf and Wang, Lina},
  date = {2006},
  journaltitle = {Journal of Computational and Applied Mathematics},
  volume = {185},
  number = {2},
  pages = {225--243},
  publisher = {Elsevier}
}

\end{document}